\documentclass[final,1p,times,twocolumn]{elsarticle}
\usepackage{epsfig}
\usepackage{amssymb}
\usepackage{amsmath}
\journal{Nuclear Physics B}

\begin{document}

\title{Effective potential of the three-dimensional Ising model:
the pseudo-$\epsilon$ expansion study}

\author[label1,label2]{A.\,I.\,Sokolov \corref{cor1}}
\ead{ais2002@mail.ru} \cortext[cor1]{Corresponding author}
\author[label1,label2]{ A.\,Kudlis}
\author[label1]{ M.\,A.\,Nikitina}

\address[label1]{Department of Quantum Mechanics, Saint Petersburg State
University, Ulyanovskaya 1, Petergof, Saint Petersburg 198504, Russia}
\address[label2]{ITMO University, Kronverkskii ave 49, Saint Petersburg
197101, Russia}

\date{\today}

\begin{abstract}
The ratios $R_{2k}$ of renormalized coupling constants $g_{2k}$ that enter
the effective potential and small-field equation of state acquire the
universal values at criticality. They are calculated for the
three-dimensional scalar $\lambda\phi^4$ field theory (3D Ising model)
within the pseudo-$\epsilon$ expansion approach. Pseudo-$\epsilon$
expansions for the critical values of $g_6$, $g_8$, $g_{10}$, $R_6 =
g_6/g_4^2$, $R_8 = g_8/g_4^3$ and $R_{10} = g_{10}/g_4^4$ originating from
the five-loop renormalization group (RG) series are derived.
Pseudo-$\epsilon$ expansions for the sextic coupling have rapidly
diminishing coefficients, so addressing Pad\'e approximants yields proper
numerical results. Use of Pad\'e--Borel--Leroy and conformal mapping
resummation techniques further improves the accuracy leading to the values
$R_6^* = 1.6488$ and $R_6^* = 1.6490$ which are in a brilliant agreement
with the result of advanced lattice calculations. For the octic coupling
the numerical structure of the pseudo-$\epsilon$ expansions is less
favorable. Nevertheless, the conform-Borel resummation gives $R_8^* =
0.868$, the number being close to the lattice estimate $R_8^* = 0.871$ and
compatible with the result of 3D RG analysis $R_8^* = 0.857$.
Pseudo-$\epsilon$ expansions for $R_{10}^*$ and $g_{10}^*$ are also found
to have much smaller coefficients than those of the original RG series.
They remain, however, fast growing and big enough to prevent obtaining
fair numerical estimates.
\end{abstract}


\maketitle

\section{Introduction}

The critical behavior of the systems undergoing continuous phase
transitions is characterized by a set of universal parameters including,
apart from critical exponents, renormalized effective coupling constants
$g_{2k}$ and the ratios $R_{2k} = g_{2k}/g_4^{k-1}$. These ratios enter
the small-magnetization expansion of free energy (the effective potential)
and determine, along with renormalized quartic coupling constant $g_4$,
the nonlinear susceptibilities of various orders:
\begin{equation}
F(z,m) - F(0,m) = {\frac{m^3 }{g_4}} \Biggl({\frac{z^2 }{2}} +
z^4 + R_6 z^6 + R_8 z^8 + R_{10} z^{10}... \Biggr),
\end{equation}
\begin{eqnarray}
\chi_4 &=& {\frac{\partial^3M}{\partial H^3}} \Bigg\arrowvert_{H = 0} = -
24 {\frac{\chi^2}{m^3}} g_4, \\
\chi_6 &=& {\frac{\partial^5M}{\partial H^5}} \Bigg\arrowvert_{H = 0} =
- 6! {\frac{\chi^3 g_4^2}{m^6}}(R_6 - 8), \\
\chi_8 &=& {\frac{\partial^7M}{\partial H^7}} \Bigg\arrowvert_{H
= 0} =
- 8! {\frac{\chi^4 g_4^3}{m^9}}(R_8 - 24 R_6 + 96), \\
\chi_{10} &=& {\frac{\partial^9M}{\partial H^9}} \Bigg\arrowvert_{H = 0} =
- 10! {\frac{\chi^5 g_4^4}{m^{12}}}(R_{10} - 32 R_8 - 18 R_6^2 + 528 R_6 - 1408),
\end{eqnarray}
where $z = M \sqrt{g_4/m^{1 + \eta}}$ is a dimensionless magnetization,
renormalized mass $m \sim (T - T_c)^\nu$ being the inverse correlation
length, $\chi$ is a linear susceptibility while $\chi_4$, $\chi_6$,
$\chi_8$ and $\chi_{10}$ are nonlinear susceptibilities of fourth, sixth,
eighth and tenth orders.

For the three-dimensional (3D) Ising model, the effective potential and
nonlinear susceptibilities are intensively studied during several decades.
In particular, renormalized coupling constants $g_{2k}$ and the ratios
$R_{2k}$ were evaluated by a number of analytical and numerical methods
\cite{B77,LGZ77,BNM78,LGZ80,BB,B1,B2,TW,B3,Reisz95,AS95,CPRV96,
S96,ZLF,SOU,GZ97,Morr,BC97,GZJ98,SO98,BC98,PV98,PV,S98,SOUK99,CPRV99,PV2000,
ZJ01,CHPV2001,CPRV2002,ZJ,PV02,CSS07,BP11,NS14}. Estimating universal
critical values of $g_4$, $g_6$ and $R_6$ by means of field-theoretical
renormalization group (RG) approach in fixed dimensions has shown that RG
technique enables one to get accurate numerical estimates for these
quantities. For example, four- and five-loop RG expansions resummed by
means of Borel-transformation-based procedures lead to the values for
$g_6^{*}$ differing from each other by less than $0.5\%$ \cite{SOU, GZ97}
while the three-loop RG approximation turns out to be sufficient to
provide an apparent accuracy no worse than $1.6\%$ \cite{SOU, S98}. In
principle, this is not surprising since the field-theoretical RG approach
proved to be highly efficient when used to estimate critical exponents,
critical amplitude ratios, marginal dimensionality of the order parameter,
etc. for numerous phase transition models
\cite{BNM78,LGZ80,AS95,GZJ98,SO98,ZJ01,ZJ,PV02,PS2000,CPV2000,PS2001}.

To obtain proper numerical estimates from diverging RG expansions the
resummation procedures have to be applied. Most of those being used today
employ Borel transformation which avoids factorial growth of higher-order
coefficients and enables one to construct converging iteration schemes.
This transformation has paved the way to a great number of high precision
numerical estimates. There exists, however, alternative technique turning
divergent perturbative series into more suitable ones, i. e. into
expansions that have smaller lower-order coefficients and much slower
growing higher-order ones than those of original series. We mean the
method of pseudo-$\epsilon$ expansion invented by B. Nickel (see Ref. 19
in the paper of Le Guillou and Zinn-Justin \cite{LGZ80}). The
pseudo-$\epsilon$ expansion approach has been shown to be very efficient
when used to estimate critical exponents and other universal quantities
for various 3D and 2D systems \cite{LGZ80,NS14,FH97,FH99,FHY2000,HDY01,
DHY02,CPN04,CP04,COPS04,HID04,DHY04,CP05,S2005,S2013,NS13,NS14e,NS16,
NS16h,KS16,KS17}.

In this paper, we study renormalized effective coupling constants and
universal ratios $R_{2k}$ of the three-dimensional Ising model with the
help of pseudo-$\epsilon$ expansion technique. The pseudo-$\epsilon$
expansions ($\tau$-series) for renormalized coupling constants $g_6$,
$g_8$ and $g_{10}$ will be calculated on the base of five-loop RG
expansions obtained by R. Guida and J. Zinn-Justin \cite{GZ97} for scalar
field theory of $\lambda\varphi^4$ type. Along with the higher-order
couplings, universal critical values of ratios $R_6 = g_6/g_4^2$, $R_8 =
g_8/g_4^3$ and $R_{10} = g_{10}/g_4^4$ will be found as series in $\tau$
up to $\tau^5$ terms. The pseudo-$\epsilon$ expansions obtained will be
processed by means of Pad\'e, Pad\'e--Borel--Leroy and conformal mapping
resummation techniques as well as by direct summation when it looks
reasonable. The numerical estimates for the universal ratios will be
compared with numerous results deduced from the higher-order
$\epsilon$-expansions, perturbative RG expansions in physical dimensions
and extracted from the lattice calculations and some conclusions
concerning the numerical power of the pseudo-$\epsilon$ expansion approach
will be formulated.

The paper is organized as follows. In the next section the
pseudo-$\epsilon$ expansions for $g_6^*$, $g_8^*$, $g_{10}^*$, $R_6^*$,
$R_8^*$ and $R_{10}^*$ are derived from 3D RG series and known
$\tau$-series for the Wilson fixed point location. Section III contains
numerical estimates for the sextic coupling resulting from the
pseudo-$\epsilon$ expansion for $R_6^*$. Sections IV deals with the
renormalized octic coupling and numerical estimates for $R_8^*$ obtained
within various resummation techniques are presented here. In Section V the
tenth-order coupling constant and the ratio $R_{10}^*$ are evaluated and
relevant numerical results are discussed. The last section contains a
summary of the results obtained.

\section{Pseudo-$\epsilon$ expansions for higher-order coupling constants}

The critical behavior of 3D Ising model is described by Euclidean field
theory with the Hamiltonian:
\begin{equation}
\label{eq:6}
H = {1\over 2}\int d^{3}x\Biggl(m_0^2 \varphi^2 + \nabla
\varphi^2 + {\lambda \over 12} \varphi^4 \Biggr),
\end{equation}
where $\varphi$ is a real scalar field, bare mass squared $m_0^2$ being
proportional to $T - T_c^{(0)}$, $T_c^{(0)}$ -- mean field transition
temperature. The $\beta$-function for the model (6) has been calculated
within the massive theory \cite{BNM78} with the propagator, quartic vertex
and $\varphi^2$ insertion normalized in a conventional way:
\begin{eqnarray}
\label{eq:7} G_R^{-1} (0, m, g_4) = m^2 , \qquad \quad {{\partial G_R^{-1}
(p, m, g_4)} \over {\partial p^2}}
\bigg\arrowvert_{p^2 = 0} = 1 , \\
\nonumber
\Gamma_R (0, 0, 0, m, g) = m^2 g_4, \qquad \quad
\Gamma_R^{1,2} (0, 0, m, g_4) = 1.
\end{eqnarray}
Later, the five-loop RG series for renormalized coupling constants $g_6$,
$g_8$ and $g_{10}$ of this model were obtained \cite{GZ97} and the
six-loop pseudo-$\epsilon$ expansion for the Wilson fixed point location
was reported \cite{NS14}. The series mentioned are:
\begin{eqnarray}
\label{eq:8}
g_6 = {\frac 9\pi }g_4^3\Bigl(1-{\frac{3}{\pi}} g_4 +
1.38996295 g_4^2 - 2.50173246 g_4^3 + 5.275903\ g_4^4 \Bigr),
\end{eqnarray}
\begin{eqnarray}
\label{eq:9}
g_8 = -{\frac{81}{{2\pi}}} g_4^4 \Bigl(1 - {\frac{65}{6\pi}}
g_4 + 7.77500131  g_4^2 - 18.5837685 g_4^3 + 48.16781 g_4^4\Bigr),
\end{eqnarray}
\begin{eqnarray}
\label{eq:10}
g_{10} = {\frac{243}{{\pi}}} g_4^5 \Bigl(1 -
{\frac{20}{\pi}} g_4 + 23.1841758 g_4^2 - 74.2747105 g_4^3 + 238.6138
g_4^4\Bigr),
\end{eqnarray}
\begin{eqnarray}
\label{eq:11}
g_4^* &=&\frac{2\pi}{9}\biggl(\tau + 0.4224965707\tau^{2} +
0.005937107 \tau^{3} + 0.011983594 \tau^{4}
\nonumber\\
&-& 0.04123101\tau^{5} + 0.0401346\tau^{6}\biggr).
\end{eqnarray}
Combining these expansions one can easily arrive to the $\tau$-series for
the higher-order coupling constants at criticality. They are as follows:
\begin{eqnarray}
\label{eq:12}
g_6^* = {8 \pi^2 \over 81}\tau^3 \bigl(1 + 0.600823045
\tau + 0.104114939 \tau^2 - 0.023565414 \tau^3 - 0.01838783 \tau^4 \bigr)
\end{eqnarray}
\begin{eqnarray}
\label{eq:13}
g_8^* = -{8 \pi^3 \over 81}\tau^4 \bigl(1 - 0.717421125 \tau
- 0.201396988 \tau^2 - 0.70623903 \tau^3 + 0.8824349 \tau^4 \bigr)
\end{eqnarray}
\begin{eqnarray}
\label{eq:14}
g_{10}^* = {32 \pi^4 \over 243}\tau^5 \bigl(1 - 2.33196159
\tau + 1.84782991 \tau^2 - 3.0485209 \tau^3 + 6.816764 \tau^4 \bigr).
\end{eqnarray}
Corresponding pseudo-$\epsilon$ expansions for the universal ratios
$R_{2k}$ read:
\begin{eqnarray}
\label{eq:15}
R_6^* = 2\tau \bigl(1- 0.244170096 \tau + 0.120059430
\tau^2 - 0.1075143 \tau^3 + 0.1289821 \tau^4\bigr).
\end{eqnarray}
\begin{eqnarray}
\label{eq:16}
R_8^* = -9\tau \bigl(1 - 1.98491084 \tau + 1.76113570
\tau^2 - 1.9665851 \tau^3 + 2.741546 \tau^4 \bigr).
\end{eqnarray}
\begin{eqnarray}
\label{eq:17}
R_{10}^* = 54\tau \bigl(1 - 4.02194787 \tau + 7.55009811
\tau^2 - 11.784685 \tau^3 + 20.05363 \tau^4 \bigr).
\end{eqnarray}
These $\tau$-series will be used for evaluation of higher-order effective
couplings near the critical point.

\section{Sextic effective interaction at criticality}

In this Section we find numerical estimates for the critical asymptote of
the ratio $R_6$ from the pseudo-$\epsilon$ expansion obtained. Since the
series (15) has small higher coefficients direct summation of this series
looks more or less reasonable. Within third, fourth and fifth orders in
$\tau$ it gives 1.752, 1.537 and 1.795 respectively, i. e. the numbers
grouping around the estimates 1.644 and 1.649 extracted from advanced
field-theoretical and lattice calculations \cite{GZ97,BP11}. It is
interesting that the value 1.537 obtained by truncation of the series (15)
by the smallest term (optimal truncation \cite{NS13}) differs from the
estimates just mentioned by 6\% only. Moreover, direct summation of
$\tau$-series for $g_6^*$ (12) having very small higher-order coefficients
gives the value $g_6^* = 1.621$ which under $g_4^* = 0.9886$ \cite{BNM78}
results in the estimate $R_6^* = 1.659$ looking rather optimistic. These
fact confirms the conclusion that the pseudo-$\epsilon$ expansion itself
may be considered as a resummation method
\cite{NS14,NS13,NS14e,NS16,NS16h}.

Much more accurate numerical value of $R_6^*$ can be obtained from the
pseudo-$\epsilon$ expansion (15) using Pad\'e approximants [L/M]. Pad\'e
triangle for $R_6^*/\tau$, i. e. with the insignificant factor $\tau$
neglected is presented in Table I.
\begin{table}[t]
\caption{Pad\'e table for pseudo-$\epsilon$ expansion of the ratio
$R_6^*$. Pad\'e approximants [L/M] are derived for $R_6^*/\tau$, i. e.
with factor $\tau$ omitted. The lowest line (RoC) shows the rate of
convergence of Pad\'e estimates to the asymptotic value. Here the Pad\'e
estimate of $k$-th order is that given by diagonal approximant or by the
average over two near-diagonal ones when corresponding diagonal
approximant does not exist.}
\label{tab1}
\renewcommand{\tabcolsep}{0.4cm}
\begin{tabular}{{c}|*{5}{c}}
$M \setminus L$ & 0 & 1 & 2 & 3 & 4  \\
\hline
0 & 2      & 1.5117 & 1.7518 & 1.5368 & 1.7947 \\
1 & 1.6075 & 1.6726 & 1.6383 & 1.6540  \\
2 & 1.6896 & 1.6465 & 1.6502 &  \\
3 & 1.6036 & 1.6504  \\
4 & 1.7135 \\
\hline
RoC & 2    & 1.5596 & 1.6726 & 1.6424 & 1.6502 \\
\end{tabular}
\end{table}
Along with the numerical values given by various Pad\'e approximants the
rate of convergence of Pad\'e estimates to the asymptotic value is shown
in this Table (the lowest line). Since the diagonal and near-diagonal
Pad\'e approximants are known to possess the best approximating properties
the Pad\'e estimate of $k$-th order is accepted to be given by the
diagonal approximant or by the average over two near-diagonal ones when
corresponding diagonal approximant does not exist. As seen from Table I,
the convergence of Pad\'e estimates is well pronounced and the asymptotic
value $R_6^* = 1.6502$ is close to the 3D RG estimate $R_6^* = 1.644 \pm
0.006$ \cite{GZ97} and, in particular, to the value $R_6^* = 1.649 \pm
0.002$ given by the advanced lattice calculations \cite{BP11}.

Let us apply further the more powerful, Pad\'e--Borel--Leroy resummation
technique. It employes the Borel--Leroy transformation of the original
diverging series
\begin{eqnarray}
\label{eq:18}
f(x) = \sum_{i = 0}^{\infty} c_i x^i = \int\limits_0^{\infty}
e^{-t} t^b F(xt) dt \ ,   \qquad \quad
F(y) = \sum_{i = 0}^{\infty} {c_i \over \Gamma(i + b + 1)} y^i =
\sum_{i = 0}^{\infty} F_i y^i  \ ,
\end{eqnarray}
with subsequent analytical continuation of the Borel-Leroy transform
$F(y)$ with a help of Pad\'e approximants. The shift parameter $b$ is
commonly used for optimization of the resummation procedure. To resum the
series (15) we address all the non-trivial Pad\'e approximants apart from
the approximant [1/4] which turned to be spoilt by a positive axis pole
for the relevant values of $b$. The Pad\'e--Borel--Leroy estimates of
$R_6^*$ given by approximants [4/1], [3/2], and [2/3] under various $b$
are listed in Table II. Since near-diagonal approximants [3/2] and [2/3]
are expected to lead to the most reliable results we concentrate on the
last two columns of this Table. It is seen that the estimates these
approximants result in are very close at any $b$. Moreover, under $b = 4$
these estimates coincide -- the curves $R_6^*(b)_{[3/2]}$ and
$R_6^*(b)_{[2/3]}$ touch (do not cross) each other at this point. So, the
value $R_6^* = 1.6488$ corresponding to this touch point may be considered
as a final Pad\'e--Borel--Leroy estimate for the effective sextic coupling
at criticality. This estimate is remarkably close to the value $1.649 \pm
0.002$ obtained recently in the course of comprehensive lattice
calculations \cite{BP11}.
\begin{table}[t]
\caption{Pad\'e-Borel-Leroy estimates of $R_6^*$ resulting from
pseudo-$\epsilon$ expansion (15) as functions of the shift parameter $b$.
Pad\'e approximants [4/1], [3/2], [2/3] are used for analytical
continuation of the Borel-Leroy transform while approximant [1/4] is
spoilt by positive axis pole for all relevant values of $b$. It is amazing
that under $b = 4$ the estimates provided by approximants [3/2] and [2/3]
coincide, i. e. the curves $R_6^*(b)_{[3/2]}$ and $R_6^*(b)_{[2/3]}$ touch
(do not cross) each other at this point.} \label{tab2}
\renewcommand{\tabcolsep}{0.4cm}
\begin{tabular}{{c}|*{3}{c}}
$b$ & [4/1] & [3/2] & [2/3] \\
\hline
0  & 1.6485 & 1.64786     & 1.64819      \\
1  & 1.6494 & 1.64822     & 1.64832      \\
2  & 1.6500 & 1.648475    & 1.648506     \\
3  & 1.6505 & 1.648672    & 1.648678     \\
4  & 1.6509 & 1.6488294   & 1.6488294    \\
5  & 1.6512 & 1.648958    & 1.648961     \\
6  & 1.6514 & 1.649065    & 1.649074     \\
7  & 1.6516 & 1.649155    & 1.649173     \\
10 & 1.6521 & 1.64936     & 1.64940      \\
15 & 1.6526 & 1.64957     & 1.64964      \\
\end{tabular}
\end{table}

What may be referred to as a measure of accuracy of the numerical result
just obtained? The choice for such a measure which looks natural is a
variation of the most stable Pad\'e--Borel--Leroy estimate --
$R_6^*(b)_{[2/3]}$ -- under $b$ varying within the whole range $[0,
\infty)$. For $b$ running from 0 to infinity $R_6^*(b)_{[2/3]}$ grows from
1.6482 to 1.6502. Hence, we adopt
\begin{eqnarray}
R_6^* = 1.6488 \pm 0.0014.
\end{eqnarray}
The inaccuracy bar accepted, being really small, is, in fact, rather
conservative since it covers as well a range of variation of the less
stable Pad\'e--Borel--Leroy estimate $R_6^*(b)_{[3/2]}$.

Although the estimate just obtained looks quite satisfactory, to confirm
its reliability and accuracy it is reasonable to resum the series (15)
employing some alternative procedure for the analytical continuation of
the Borel--Leroy transform. We address here the conformal mapping
technique which, being widely used in the theory of critical phenomena, is
known to demonstrate high numerical efficiency. This technique is based on
the knowledge of the closest singularity of the Borel--Leroy transform
(\ref{eq:18}) from the large-order behavior of the original series
coefficients. For the Borel--Leroy image $F(y)$ it assumes an analyticity
in the complex $y$-plane with a cut from $-1/\alpha$ to $-\infty$. Since
the series for $F(y)$ converges within the circle $|y| < 1/\alpha$ the
integration of (\ref{eq:18}) requires its analytical continuation. It is
performed by means of a conformal mapping $y = f(w)$ that should transform
the plane with a cut into the unit circle $|w| < 1$ and shift the
singularities lying on the negative real axis on the circumference of the
circle \cite{LGZ77}. The conformal mapping
\begin{equation}
w(y)=\frac{\sqrt{1+\alpha y}-1}{\sqrt{1+\alpha y}+1} \ , \qquad \quad
y(w)=\frac{4}{\alpha}\frac{w}{(1-w)^2} \
\end{equation}
is easily seen to satisfy the aforementioned requirements. Thus,
we arrive to a new series:
\begin{equation}
F(y(w))=\sum\limits_{i=0}^{\infty}F_i\left(\frac{4}{\alpha}\right)^i\frac{w^i}{(1-w)^{2i}} =
\sum\limits_{i=0}^{\infty}W_i w^i(y) \ ,
\end{equation}
\begin{equation}
W_0 = B_0 \ , \quad \qquad W_n =
\sum\limits_{i=1}^{n}F_i\left(\frac{4}{\alpha}\right)^i\frac{(i+n-1)!}{(n-i)!(2i-1)!}
\end{equation}
that converges for any $y$. In our case only first $N$ coefficients of the
expansion are known, i. e. we have to work with the truncated series for
Borel--Leroy transform:
\begin{equation}
F^{(N)}(y)=\sum\limits_{i=0}^{N}F_i y^i = \sum\limits_{i=0}^{N}W_i w^i(y).
\end{equation}
Correspondingly, the conform-Borel-resummed series for $f(x)$ is given by
the following expression:
\begin{equation}
f^{(N)}(y)=\sum\limits_{i=0}^{N}W_i\int\limits_{0}^{\infty} e^{-t} t^b
\Biggl(\frac{\sqrt{1+\alpha y t }-1}{\sqrt{1+\alpha y t}+1}\Biggr)^i dt.
\end{equation}

For 3D Ising model the constant $\alpha$ was evaluated in a course of the
large-order perturbative analysis \cite{ZJ}:
\begin{equation}
\alpha = 0.147774232.
\end{equation}
In principle, parameter $b$ for various universal quantities may be taken
from the same source \cite{BGZ77}. On the other hand, the five-loop
approximation is obviously can not be thought of as lengthy enough to
match the Lipatov's asymptotics. That is why it looks natural to consider
$b$ as a fitting parameter which may be used for optimization of the
resummation procedure. We mean here the acceleration of the series
convergence and, especially, the stability of the numerical estimate under
the variation of the fitting parameter.

\begin{figure}[h!]
\centering
\includegraphics[scale=0.3]{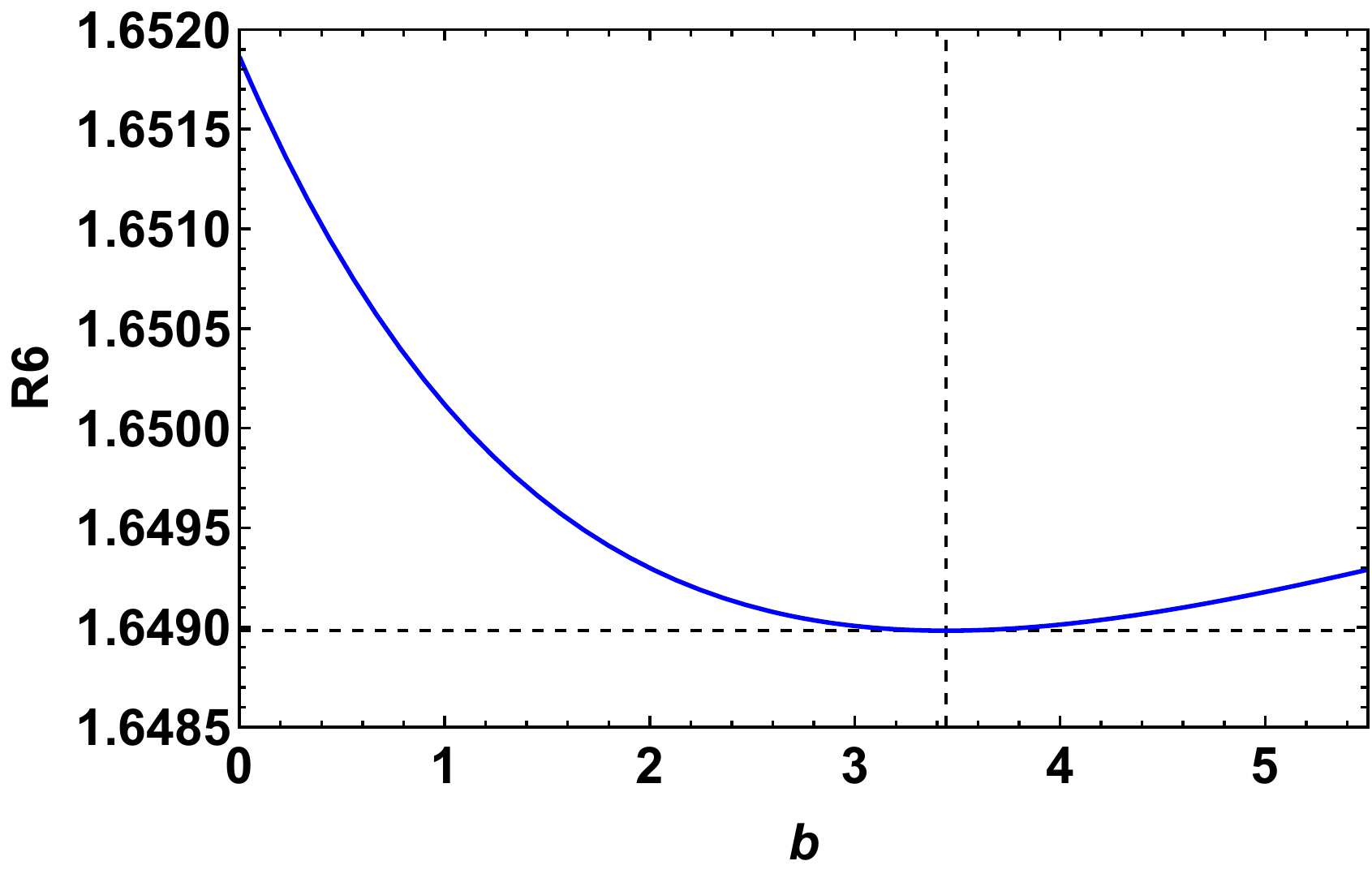}
\caption{The value of $R_6^*$ as a function of the parameter $b$ obtained
by means of conformal mapping technique.}
\label{fig-1}
\end{figure}

In Fig. \ref{fig-1} the
numerical value of $R_6^*$ given by conform-Borel resummed series (15) is
shown as a function of $b$. As is seen, the region where the universal
ratio demonstrates a minimal sensitivity (maximal stability) with respect
to $b$ is centered near $b \approx 3.5$. More precisely, the curve
$R_6^*(b)$ has an extremum at $b = 3.4434$ which corresponds to $R_6^* =
1.64898$. Thus we adopt the value
\begin{equation}
R_6^* =1.6490
\end{equation}
as a final estimate the conformal mapping technique yields. This number is
seen to be in a complete agreement with results obtained by 3D RG analysis
and advanced lattice calculations.

It is worthy to note that an account for the five-loop terms in the RG
expansion and $\tau$-series for $R_6^*$ shifts the numerical value of this
universal ratio only slightly. Indeed, the Pad\'e--Borel--Leroy
resummation of the four-loop RG series for $g_6$ leads to $R_6^* = 1.648$
\cite{SOUK99} while Pad\'e resummed four-loop $\tau$-series for $R_6^*$
and $g_6$ result in $R_6^* = 1.642$ and $R_6^* = 1.654$, respectively
\cite{NS14}. This may be considered as an extra manifestation of the fact
that the RG expansion and $\tau$-series for $R_6^*$ have a structure
rather favorable from the numerical point of view. It is especially true
for the pseudo-$\epsilon$ expansion (15) which being alternating and
having small higher-order coefficients turns out to be very convenient for
getting numerical estimates.

As was recently shown \cite{NS14}, the pseudo-$\epsilon$ expansions of
$R_6^*$ for the systems with $n$-vector order parameter (easy-plane and
Heisenberg ferromagnets, etc.) have smaller higher-order coefficients than
those for the Ising ($n = 1$) model. This implies that for $n > 1$ the
iteration procedure based on the Pad\'e--Borel--Leroy resummation
technique should converge faster and give better numerical results than
for $n = 1$. So, we believe that for $n > 1$ the four-loop
pseudo-$\epsilon$ expansions \cite{NS14} will give the numerical estimates
of $R_6^*$ practically as precise as that given by (unknown) five-loop
$\tau$-series provided the Pad\'e--Borel--Leroy resummation is made. Here
we present such four-loop Pad\'e--Borel--Leroy estimates for $XY$ ($n =
2$) and Heisenberg ($n = 3$) models, i. e. for the systems most
interesting from the physical point of view:
\begin{eqnarray}
R_6^* = 1.570 \pm 0.007 \quad(n = 2), \qquad  R_6^* = 1.500 \pm
0.005 \quad(n = 3).
\end{eqnarray}
These numbers are in a good agreement with other field-theoretical and
lattice estimates \cite{SOUK99,PV2000,NS14,CHPRV2001,CHPRV2002}.

\section{Octic coupling: resummation and numerical estimates}

As seen from Eqs. (\ref{eq:13}) and (\ref{eq:16}), for the renormalized
octic coupling we have pseudo-$\epsilon$ expansions with less favorable
structure. The series for $R_8^*$ being alternating have big higher-order
coefficients. To estimate this ratio we'll apply various resummation
procedures -- the Pad\'e, Pad\'e--Borel--Leroy and conformal mapping
techniques. Higher-order coefficients of the $\tau$-expansion for $g_8^*$
are much smaller but have irregular signes. This series will be also
processed within the techniques mentioned aiming to find the universal
value of $R_8$ via the relation $R_8 = g_8/g_4^3$.

Let us start estimating the universal value of the octic coupling from the
Pad\'e approximant approach. Pad\'e triangle for pseudo-$\epsilon$
expansion of the ratio $R_8^*/\tau$, i. e. with the insignificant factor
$\tau$ omitted is presented in Table \ref{tab3}. The rate of convergence
of Pad\'e estimates to the asymptotic value is also shown at the bottom of
this table. As one can easily see the convergence of the Pad\'e estimates
for $R_8^*$ is much less pronounced than in the case of $R_6^*$. At the
same time, the simple method employed gives the asymptotic value $R_8^* =
0.879$ that is in a good agreement with the result of lattice calculations
$R_8^* = 0.871\pm 0.014$ \cite{BP11} and with the number $R_8^*=0.857 \pm
0.086$ given by the 3D RG analysis \cite{GZ97}.
\begin{table}[t]
\caption{Pad\'e triangle for pseudo-$\epsilon$ expansion of the ratio
$R_8^*$. Pad\'e approximants [L/M] are constructed for $R_8^*/\tau$, i. e.
with factor $\tau$ omitted. The lowest line (RoC) demonstrates the rate of
convergence of Pad\'e estimates where Pad\'e estimate of $k$-th order is
that given by diagonal approximant or by the average over two
near-diagonal ones if diagonal approximant does not exist.}
\label{tab3}
\renewcommand{\tabcolsep}{0.4cm}
\begin{tabular}{{c}|*{5}{c}}
$M \setminus L$ & 0 & 1 & 2 & 3 & 4 \\
\hline
0   & $-9$     & 8.864  & $-6.986$ & 10.713   & $-13.961$ \\
1   & $-3.015$ & 0.466  & 1.376    & 0.407 \\
2   & $-1.743$ & 1.910  & 0.879    \\
3   & $-1.131$ & 0.095  \\
4   & $-0.831$ \\
\hline
RoC & $-9$     & 2.925  & 0.466    &  1.643   & 0.879 \\
\end{tabular}
\end{table}

The next method of the resummation which we will address is the
Pad\'e--Borel--Leroy technique. This technique reduces to the
Pad\'e--Borel resummation procedure when one put the fitting parameter $b$
equal to zero. Let us first present the estimates of $R_8^*$ that this
simpler machinery yields. They are collected in Table \ref{tab4}. As is
seen, use of the Borel transformation significantly accelerates the
convergence of numerical estimates to the asymptotic value, but this value
itself -- $R_8^* = 0.890$ -- turns out to be slightly further from the
results of 3D RG analysis and advanced lattice calculations than its
Pad\'e counterpart.
\begin{table}[t] \label{tab4}
\caption{Pad\'e-Borel estimates extracted from the pseudo-$\epsilon$
expansion for $R_8^*$. Pad\'e approximants [L/M] are used for analytical
continuation of the Borel transform. The lowest line (RoC) shows the rate
of convergence of Pad\'e-Borel estimates to the asymptotic value.
Pad\'e-Borel estimate of $k$-th order is that originating from diagonal
approximant for $R_8^*/\tau $ or by average over two near-diagonal ones
when corresponding diagonal approximant does not exist.}
\label{tab4}
\renewcommand{\tabcolsep}{0.4cm}
\begin{tabular}{{c}|*{5}{c}}
$M \setminus L$ & 1 & 2 & 3 & 4 & 5 \\
\hline
0 & $-9$      & 8.864   & $-6.986$ & 10.713  & $-13.961$ \\
1 & $-3.6473$ & 1.1000  & 0.88541  & 0.89064 \\
2 & $-2.4658$ & 0.90410 & 0.89048  \\
3 & $-2.0048$ & 0.89164 \\
4 & $-1.7788$ \\
\hline
RoC &  $-9$   & 2.608   &  1.1000  & 0.8948  & 0.89048 \\
\end{tabular}
\end{table}

\begin{figure}[h!]
\centering
\includegraphics[scale=0.3]{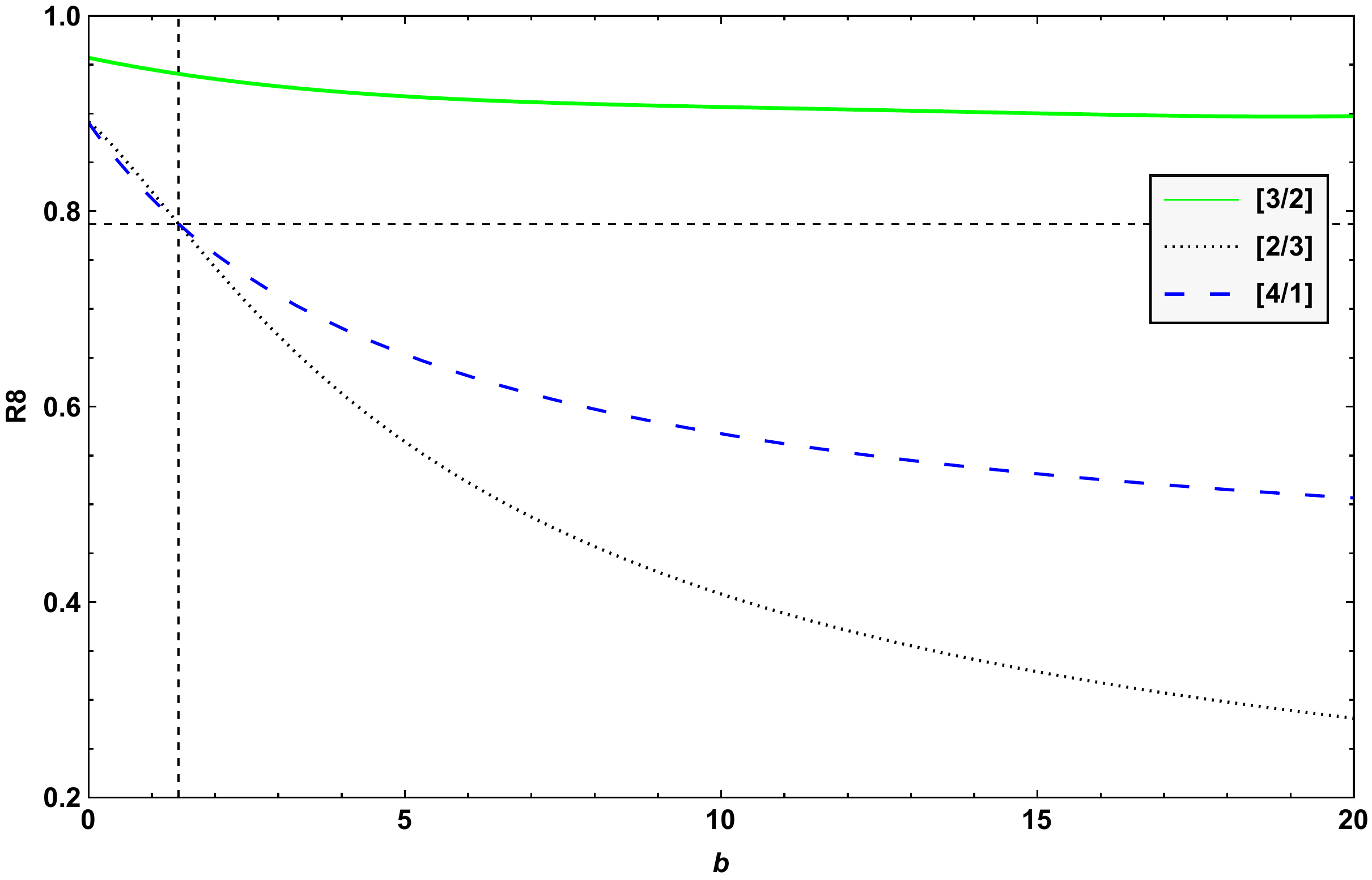}
\caption{Pad\'e--Borel--Leroy estimates of the universal ratio $R_8^*$
based upon approximants [3/2], [2/3] and [4/1] as functions of the
parameter $b$.} \label{fig-2}
\end{figure}

Can the situation be improved by making the fitting parameter $b$ active?
To clear up this point, we sum up the series (\ref{eq:16}) by means of the
Pad\'e--Borel--Leroy technique using as working three Pad\'e approximants
-- [3/2], [2/3] and [4/1]. The results are presented at Fig. \ref{fig-2}
where corresponding estimates as functions of $b$ are shown.

One can see from this figure that the pseudo-$\epsilon$ expansion
(\ref{eq:16}) has much worse approximating properties than that of the
$\tau$-series for $R_6^*$. Indeed, only one approximant -- [3/2] -- gives
the estimates that are stable with respect to the variation of $b$. They
are grouped around the number 0.9 that therefore may be thought of as
close to the true value. On the other hand, working approximants [2/3] and
[4/1] result in the different estimate: curves $R_8^*(b)_{[2/3]}$ and
$R_8^*(b)_{[4/1]}$ intersect under $b \approx 1.5$ thus yielding $R_8^* =
0.8$, the value which also looks plausible. Since the numbers just found
disagree with each other by more than 10$\%$ none of them can be
considered as reliable.

In such situation an alternative resummation procedure should be applied.
As before, we use the conformal mapping technique. The results of the
resummation of series (\ref{eq:16}) by means of conform-Borel approach
under varying $b$ are presented in Fig. 3.
\begin{figure}[h!]
\centering
\includegraphics[scale=0.3]{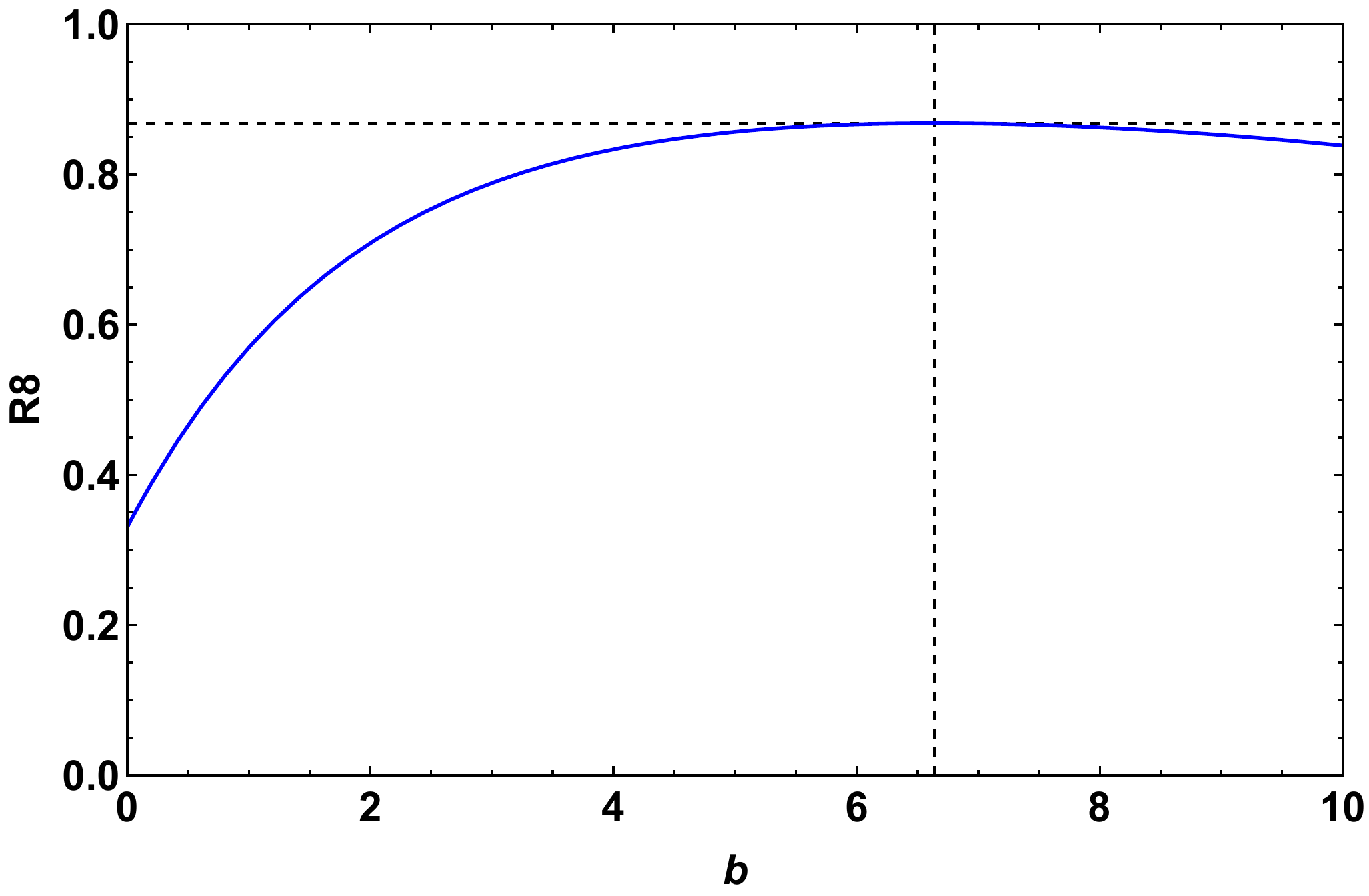}
\caption{The value of $R_8^*$ obtained by means of conformal mapping
technique as function of the parameter $b$.}
\label{fig-3}
\end{figure}
The curve $R_8^*(b)_{cB}$ is seen to have a smooth extremum and,
correspondingly, a wide enough region where the numerical estimate for
$R_8^*$ is stable with respect to $b$. The value of octic coupling at the
extremum $b = 6.638$, i. e. at the point of maximal stability is
\begin{equation}
R^*_8 = 0.868.
\end{equation}

As was already noted, the universal ratio $R^*_8$ may be also found via
evaluation of the octic coupling constant $g_8$ at criticality and use of
the relation $R_8 = g_8/g_4^3$. Since the coefficients of the
pseudo-$\epsilon$ expansion (\ref{eq:13}) are considerably smaller than
those of the $\tau$-series for $R^*_8$ this way seems to be promising.
However, the expansion (\ref{eq:13}) has rather irregular structure and
all attempts to sum up this series with a help of above methods have
failed to yield satisfactory results.

So, we accept the number (28) as a final result of our pseudo-$\epsilon$
expansion analysis. It turns out to be close to the value $R_8^* =
0.871\pm 0.014$ extracted from the most recent lattice calculations
\cite{BP11} and compatible with the 3D RG estimate $R_8^*=0.857 \pm 0.086$
\cite{GZ97}.

\section{Universal ratio $R_{10}$}

Comparing original RG expansion (10) with the series (14) and (17) one can
see that the application of the pseudo-$\epsilon$ expansion technique
certainly improves the structure of the series for $g^*_{10}$ and
$R^*_{10}$: it significantly diminishes the coefficients leaving the
series alternating. At the same time, the coefficients of the
pseudo-$\epsilon$ expansions remain big and fast growing what makes direct
or Pad\'e summation of the series (14) and (17) meaningless. Attempting to
arrive to proper numerical estimates we sum up the series (17) with a help
of the Pad\'e--Borel--Leroy procedure. The results thus obtained are shown
in Fig. 4. As is seen, the values of $R^*_{10}$ given by four relevant
approximants strongly depend on the shift parameter $b$ and markedly
differ from each other. There exist, however, two points on the $b$ axis
that may be thought of as corresponding to some meaningful results.

\begin{figure}[h!]
\centering
\includegraphics[scale=0.3]{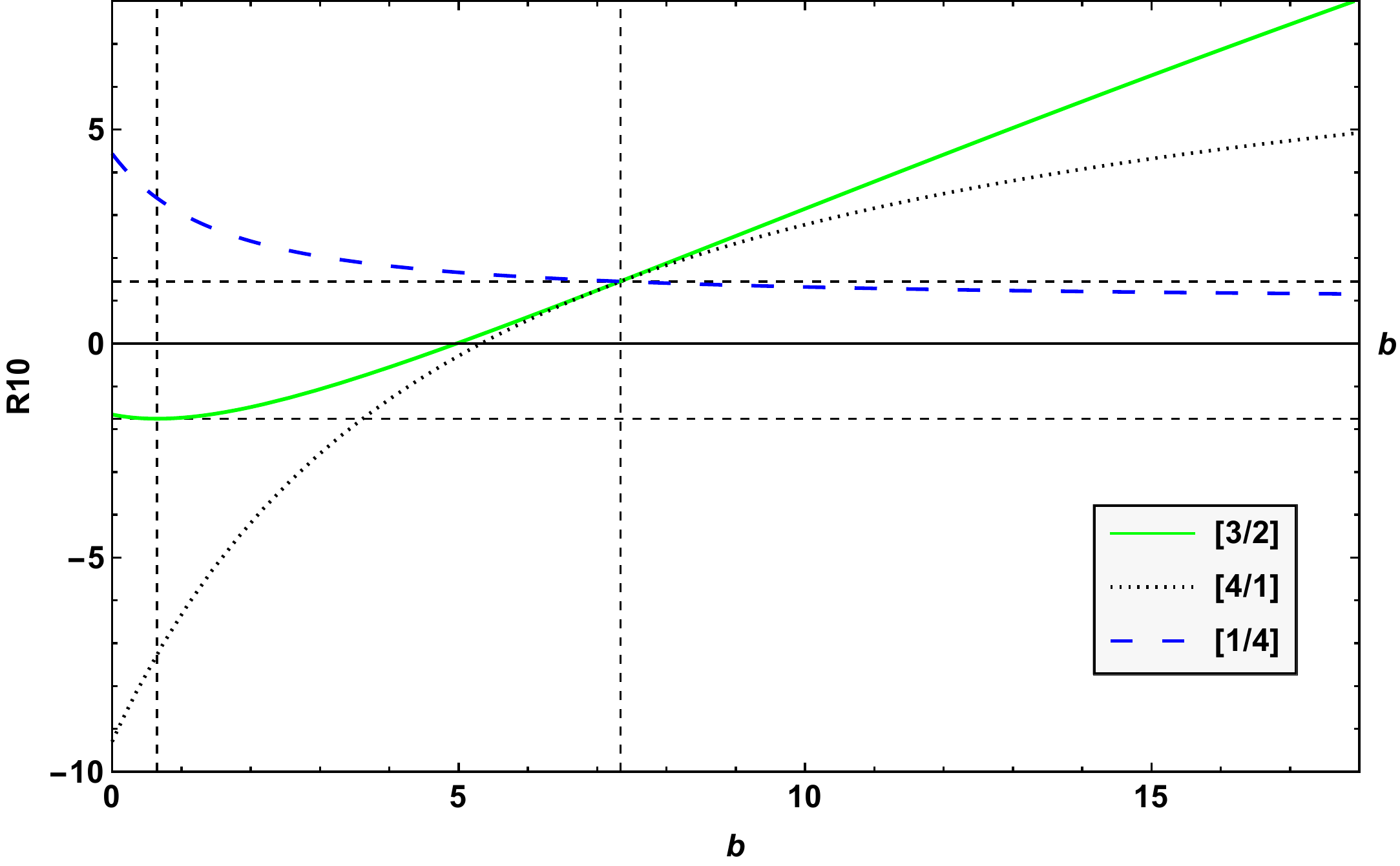}
\caption{Pad\'e--Borel--Leroy estimates of the universal ratio $R_{10}^*$
as functions of parameter $b$.}
\label{fig-3}
\end{figure}

The first one -- $b = 7.33$ -- is the point of consensus of three working
approximants, i. e. the point where use of approximants [4/1], [3/2] and
[1/4] yields practically the same value. This value $R^*_{10} = 1.45$,
however, is in an obvious disagreement with the numbers obtained by other
methods; they are collected in Table 5.
\begin{table}[t]
\caption{The values of the universal ratio $R_{10}$ obtained in this work,
found by resummation of 3D RG series and $\epsilon$-expansions, extracted
from the lattice calculations (LC) and given by the
exact-renormalization-group analysis (ERG).}\label{Table V}
\renewcommand{\tabcolsep}{0.2cm}
\begin{tabular}{|c|cccc|}
\hline This work & $ 1.45$, &  $-1.76$ &  &   \\
\hline 3D RG  & $-2.06\pm1.3$ \cite{GZ97}, & $-2.3\pm1.6$ \cite{GZJ98} &  &  \\
\hline $\epsilon$-expansion & $-1.06\pm0.1$ \cite{GZ97}, &
$-1.1\pm0.1$ \cite{GZJ98}, & $-1.8\pm1.4$ \cite{PV} &    \\
\hline LC & $-0.75\pm0.38$ \cite{BC97}, & $-1.2\pm0.4$ \cite{CPRV99}, &
$-0.97\pm0.17$ \cite{CPRV2002}, & $-1.39\pm0.04$ \cite{BP11}  \\
\hline ERG &  $-1.65\pm0.4$ \cite{Morr}, & $-1.152$ \cite{CSS07} &  &   \\
\hline
\end{tabular}
\end{table}
The second point -- $b = 0.647$ -- is the point of maximal stability of
the estimate obtained on the base of the near-diagonal approximant [3/2].
Addressing this point leads to $R^*_{10} = -1.76$, the value which is
compatible with many of the results presented in Table 5. This value,
nevertheless, can not be referred to as fair since it is in conflict with
the former result $R^*_{10} = 1.45$ obtained within the same technique.
The attempts to get reasonable estimates for $R^*_{10}$ using
conform-Borel technique or via resummation of the series (14) for the
coupling constant $g^*_{10}$ itself also turned out to be unsuccessful.

So, we see that although the pseudo-$\epsilon$ expansion machinery is able
to transform strongly divergent 3D RG expansions into series with smaller
and slower growing coefficients, in the case of $R_{10}$ it is not
powerful enough to provide acceptable numerical estimates for its
universal value.

\section{Conclusion}

To summarize, we have calculated pseudo-$\epsilon$ expansions for the
universal values of renormalized coupling constants $g_6$, $g_8$, $g_{10}$
and of the universal ratios $R_6$, $R_8$, $R_{10}$ for 3D Euclidean scalar
$\lambda\phi^4$ field theory. Numerical estimates for $R_6^*$ have been
found using Pad\'e, Pad\'e--Borel--Leroy and conform--Borel resummation
techniques. The pseudo-$\epsilon$ expansion machinery has been shown to
lead to high-precision value of $R_6^*$ which is in very good agreement
with the numbers obtained by means of other methods including advanced
lattice calculations. For the octic coupling this technique was shown to
be less efficient: numerical estimates extracted from $\tau$-series for
$R_8^*$ by means of the Pad\'e--Borel--Leroy and conform-Borel
resummations slightly differ from each other and from their lattice and 3D
RG counterparts. Corresponding differences, however, are small, especially
between conform-Borel (0.868) and advanced lattice (0.871) estimates,
indicating that pseudo-$\epsilon$ expansion technique provides precise
enough numerical results for this coupling. Pseudo-$\epsilon$ expansion
for the ratio $R_{10}^*$ has been also analyzed. It has been shown that
the pseudo-$\epsilon$ expansion approach improves the structure of series
for $g_{10}^*$ and $R_{10}^*$ but such an improvement turns out to be
insufficient to make the series suitable for getting numerical estimates.

\section*{Acknowledgment}
We gratefully acknowledge the support of Saint Petersburg State University
via Grant 11.38.636.2013, of the Russian Foundation for Basic Research
under Project No. 15-02-04687, and of the Dynasty foundation.

\end{document}